\documentclass[useAMS,usenatbib]{mn2e}
\usepackage{epsfig}
\title[Satellite galaxies of M106]{A Wide Field Survey of Satellite Galaxies around the Spiral Galaxy M106}

\author[E. Kim et al.]{E. Kim,$^1$\thanks{E-mail:
ekim@csa.yonsei.ac.kr} M. Kim,$^2$
N. Hwang,$^3$ M. G. Lee,$^4$ M.-Y. Chun,$^2$ and H. B. Ann$^5$\thanks{Author to whom any correspondence should be addressed. E-mail:hbann@pusan.ac.kr}\\
$^1$Department of Astronomy \& Center for Galaxy Evolution Research, Yonsei University, Seoul 120-749, Korea\\
$^2$Korea Astronomy and Space Science Institute, Daejeon 305-348, Korea\\
$^3$National Astronomical Observatory of Japan, 2-21-1 Osawa Mitaka, Tokyo 181-8588, Japan\\
$^4$Astronomy Program, Department of Physics and Astronomy, Seoul National University, Seoul 151-742\\
$^5$Division of Science Education, Pusan National University, Busan 609-735, Korea}

\begin{document}

\date{Accepted 20?? ?? ??. Received 2010 September 6; in original form 2010 September 6}

\pagerange{\pageref{firstpage}--\pageref{lastpage}} \pubyear{2010}

\maketitle

\label{firstpage}

\begin{abstract}
We present a wide field survey of satellite galaxies in M106 (NGC 4258)
covering a $1.7\degr \times 2\degr$ field around M106 using Canada-France-Hawaii
Telescope/MegaCam.
We find 16 satellite galaxy candidates of M106.
 Eight of these galaxies are found to be dwarf galaxies that are much 
smaller and fainter than the remaining galaxies.
Four of these galaxies are new findings.
Surface brightness profiles of 15 out of 16 satellite galaxies can be
represented well by an exponential disk profile with varying scale length.
We derive the surface number density distribution of these satellite 
galaxies.
The central number density profile (d $<100$ kpc) is 
well fitted by a power-law with a power index of $-2.1\pm0.5$, similar to the expected
power index of isothermal distribution.
The luminosity function of these satellites is represented well by the
Schechter function with a faint end slope of $-1.19^{+0.03}_{-0.06}$.
Integrated photometric properties (total luminosity, total colour, and 
disk scale length) and the spatial distribution 
of these satellite galaxies are found to be roughly 
similar to those of the Milky Way and M31.
\end{abstract}

\begin{keywords}
galaxies: individual (M106) ---galaxies: morphology and photometry --- galaxies: dwarf galaxies
\end{keywords}

\begin{table*}
 \centering
 \begin{minipage}{112mm}
  \caption{Basic information of M106 (NGC 4258)}
  \begin{tabular}{@{}ccc@{}} \hline\hline
   Parameter & Values & References \\ \hline\hline
   RA(2000), Dec(2000) & 12h18m57.5046s,  +47$^\circ$ 18$\arcmin$ 14$\arcsec.303$ & 1 \\
   Ellipticity & 0.611 & 2 \\
   P.A. & 150 deg & 2 \\
   Standard radius, $R_{25}$     & 9.3 arcmin & 2\\
   Total magnitudes              & $V^T=8.41\pm0.08$, $B^T=9.10\pm0.07$ & 2\\
   X-ray luminosity   &  $Log(L_X) = 40.88$ erg s$^{-1}$ & 3 \\
   Heliocentric radial velocity, $v_r$                 & $448\pm3$ km s$^{-1}$ & 1\\
   Foreground reddening         & $E(B-V)=0.016$  & 4\\
   Distance &  d=7.2 Mpc ($(m-M)_0=29.29\pm0.07$)  & 5,6 \\
   \hline
  \end{tabular}
  (1) NASA Extragalactic Database (NED);
  (2) RC3;
  (3) Panessa et al (2006) (nuclear flux - Chandra observation);
  (4) Schlegel et al. (1998);
  (5) Herrnstein et al. (1999);
  (6) \citet{mag08}
 \end{minipage}
\end{table*}

\section{Introduction}

Dwarf galaxies are believed to be the most abundant population in
the universe.  The preponderance of dwarf galaxies has been predicted
by cold dark matter (CDM) simulations (eg., \citet{kly99}) but the slope of
the observed luminosity function of faint galaxies is too shallow to be
matched with the predictions of CDM cosmology. Another important prediction of
hierarchical structure formation in the CDM or $\Lambda$CDM cosmology is that
most of the dwarf galaxies are thought to be satellites of giant galaxies
because galaxies are assembled by sub-structures of sub-galactic masses
\citep{fre88} and many low-mass haloes can survive after major and minor
mergers.

To date there are only a few giant spiral galaxies which are known to have
more than dozens of satellite galaxies: the Milky Way Galaxy (MW), M31 and the
M81 group. The
number of satellite galaxies in MW and M31 has been doubled recently due to
new discoveries \citep{zuc04, wil05a, wil05b, bel06, bel07a, bel07b,
gri06, sak06, zuc06a, zuc06b, zuc07, mar06, maj07, iba07, wal07}, mostly
from the Sloan Digital Sky Survey (SDSS) \citep{yor00}. Three of
them (Willman 1, Segue 1, Bootes II) are very faint ($M_{V}\sim-3$) and it is
uncertain whether they are genuine dwarf galaxies \citep{wil05b, bel07b, 
wal07}. The number of satellite galaxies in the M81 group has been 
doubled too by the
recent discoveries of 22 new satellite galaxies \citep{chi09} from a deep
survey of $8\degr \times 8\degr$ area using MegaCam of CFHT.
Among them, the satellite galaxies of the M81 group are of particular 
interest because
they are faint enough to be compared with those of MW and M31, and provide
unique information about the formation of satellite systems in a group
environment that is thought to be different from that of the Local Group.
Because of the very wide searching area of $8\degr \times 8\degr$, 
corresponding to
500 kpc$\times$500 kpc at the distance of 3.6 Mpc, \citet{chi09} 
fairly well covered all the satellite galaxies within the virial radius of M81.
The dramatic increase of the number of satellite galaxies in these systems
seems to greatly alleviate the missing satellite problem.

However, there is still a discrepancy between the number of observed
dwarf satellite galaxies and those predicted by the CDM hierarchical
models \citep{kly99, moo99}, because the number of observed MW satellites
corrected for the sky coverage of the SDSS is still a factor of 4 smaller
than the model predictions \citep{sim07}.
There have been several explanations to solve the discrepancy between
the predicted and observed numbers of dwarf galaxies. Some solutions modify
the primordial density fluctuations \citep{kam00, zen03} or the dark matter
properties by invoking warm dark matter \citep{col00, bod01}. However,
more appealing explanations employ astrophysical solutions to suppress gas
accretion on the low mass dark haloes or to prevent star formation in these
haloes so that they can not be observed as dwarf galaxies. Among others
photoionization squelching after re-ionization \citep{bul00, som02, ben02}
seems to be the most attractive because it provides a natural low mass
cutoff \citep{kop09}.
However, \citet{kra04} argued that the star formation history in the 
low mass
dwarf galaxies is not sensitive to UV background and re-ionization but
sensitive to the physics of galaxy formation. 
They proposed tidal stripping as an
alternative mechanism for the paucity of observable dwarf galaxies at
the present epoch.

Owing to the recent discoveries of new satellite galaxies in MW, M31, 
and the M81 group 
and some models that can explain the observed properties of satellite
galaxies belonging to these galaxy haloes, the missing satellite 
problem seems to be
less severe than before. However, satellite galaxies in 
the Local Group (LG) may not be a representative population of dwarf 
galaxies surrounding dark haloes because LG is located in a 
underdense region. Since the number of bright members of the M81 group is
similar to that of LG and their local background densities seem to
be not much different due to the very close distance between LG and M81.
Thus, we need more sample in different environment to understand this
problem.

\begin{figure*}
 \epsfig{figure=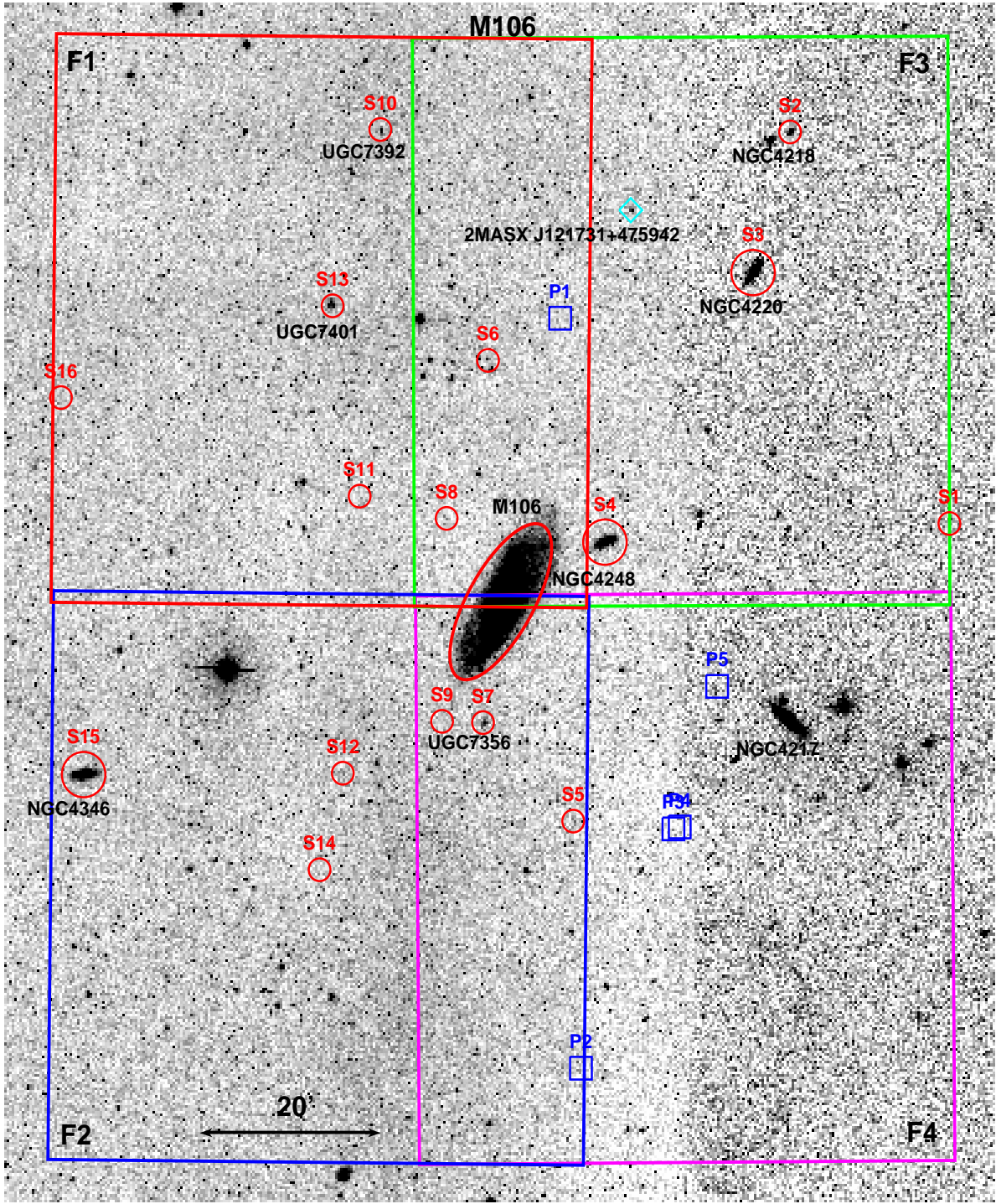, height=1.200\textwidth, width=0.990\textwidth}
 \vspace{3mm}
 \caption{
  A finding chart for the four fields of M106 marked on the Digitized Sky Survey image.
  The size of field of view is $2.2\degr \times 2.3\degr$.  North is up and
  east is to the left.
  Satellite galaxies are marked by circles with running ids
  and the extent ($D_{25}$) of M106 is shown by a large ellipse. Probable
  satellite galaxies P1$-$P5 of M106 are marked with open squares.}
 \label{fig_fov}
\end{figure*}

The purpose of the present study is to find dwarf satellite galaxies
surrounding a giant spiral galaxy which is similar to MW and M31 
to address the missing satellite problem in different
group environment.
We selected M106 (NGC 4258) as the best target for our deep and wide 
imaging survey
to search for satellite galaxies because of its
proximity ($\sim7.2$Mpc) and brightness ($M_V=-20.93$) that allow us 
to detect faint satellite galaxies down to $M_V\approx-10$.
M106 is the brightest member of CVn II group which is thought to be one of
the most dense groups in the local universe according to the 
catalogue of \citet{fou92}. \citet{tur76} assigned M106
to their Group 50 of which some galaxies are members of CVn II.

The morphology of M106 is classified as a barred spiral galaxy (SAB(s)bc).
M106 is known as a LINER 
harboring nuclear water masers which can be used for an accurate
distance estimate by geometrical means. The distance estimated from 
the water maser is $7.2\pm 0.3$Mpc \citep{her99}, which 
is consistent with the distances
determined using the Period$-$Luminosity relation of Cepheids 
and TRGB \citep{mac06,ben07,mag08,bon08}.
At this distance one arcsec corresponds to a linear scale of $35$pc.
The luminosity of M106 is slightly fainter than those of MW and M31.
Its systemic velocity is $448 \pm3$ km s$^{-1}$. 
Its maximum rotation velocity, 208 km s$^{-1}$, is slightly smaller than
that of MW \citep{eri99}. It is known to have one satellite galaxy,
NGC 4248 ($\sim13$ arcmin from the centre of M106).
Foreground reddening toward M106 is low: $A_V=0.05$, $A_R=0.04$ and $A_I=0.03$
\citep{sch98}. Basic information on M106 is summarized in Table 1.

We surveyed an area of $1.7\degr \times 2\degr$ centered on M106,
which corresponds to 215 kpc$\times$ 260 kpc, using $g$ and $r$ filters.
Our surveyed area is expected to cover about half of the virial radius
where most of the dwarf satellite galaxies are considered to reside. The
model of \citet{moo99} predicted that about 75\% of the dwarf satellite 
galaxies are located within $r=100$ kpc from the centre of the host dark
halo. Thus, we expect to detect more than 75\% of the satellite galaxies
from the present survey.

\begin{table*}
 \centering
 \begin{minipage}{145mm}
 \raggedright
  \caption{Observation log}
  \begin{tabular}{@{}ccccccccc@{}} \hline\hline
   Field & Filter & Exp.Times & Airmass & Seeing$^a$ & Date(UT) &
   Sky Brightness$^b$ & Lunar Angle$^c$ & Lunar Phase \\ \hline\hline
   F1 & g & $4 \times 210$s & 1.15 & $0\arcsec.61$ & 2005 Feb 11 & 22.42$\pm$0.05 & 133 & 0.15 \\
   F1 & r & $4 \times 600$s & 1.17 & $0\arcsec.56$ & 2005 Feb 11 & 21.58$\pm$0.03 & 133 & 0.15 \\
   F2 & g & $4 \times 210$s & 1.19 & $0\arcsec.62$ & 2005 Feb 16 & 22.19$\pm$0.06 &  95 & 0.63 \\
   F2 & r & $4 \times 600$s & 1.15 & $0\arcsec.67$ & 2005 Feb 16 & 21.50$\pm$0.03 &  94 & 0.63 \\
   F3 & g & $4 \times 210$s & 1.18 & $0\arcsec.78$ & 2005 Mar 17 & 21.26$\pm$0.05 &  79 & 0.46 \\
   F3 & r & $4 \times 600$s & 1.15 & $0\arcsec.76$ & 2005 Mar 17 & 20.94$\pm$0.02 &  78 & 0.46 \\
   F4 & g & $4 \times 210$s & 1.15 & $0\arcsec.95$ & 2005 Apr 06 & 21.77$\pm$0.02 & 140 & 0.01 \\
   F4 & r & $4 \times 600$s & 1.17 & $0\arcsec.71$ & 2005 Apr 06 & 21.11$\pm$0.02 & 140 & 0.01 \\
   \hline
  \end{tabular}
  $^a$Typical seeing size at the centre of focal plane. Significant image quality degradation
        occurs in the outer regions due to the large-format CCDs.

  $^b$In units of mag arcsec$^{-2}$.

  $^c$In units of degree.

 \end{minipage}
\end{table*}
In this paper we report on the candidate satellite galaxies of M106 some of which
are newly discovered dwarf galaxies
from a deep MegaCam survey.
Observations and data reduction are described in \S2.
We present the main results of the present study in
\S3. In \S4 we discuss the misdentifications found in the present study 
and missing satellite galaxies, and
compare the satellite system of M106 to those of LG and the M81 group.
We summarize the main results of this study in the last section.

\section{Observations and Data Reduction}

\subsection{Observations}

CCD images of M106 were obtained on the 4 nights from February 11 2005 to April 6 2005 (UT)
using the wide-field camera, MegaPrime/MegaCam at CFHT. 
Most of observations were carried out under 
a good photometric condition except for the observations of March 17 2005 when it was cloudy.
MegaCam is made of 36 $2048 \times 4612$ pixel CCDs (4 rows $\times$ 9 columns of CCDs), 
and it covers  $1\degr \times 1\degr$ field of view with a resolution of 0.185 arcsec per pixel at CFHT
MegaPrime focus.
SDSS-$g$ and $r$ filters were used for our observation. 
Four fields around M106 (F1, F2, F3, and F4) were observed and the total field coverage is
$1.7^o \times 2^o$, as displayed in Fig. 1.
All fields were observed to be partially 
overlapped with each other to check internal photometric accuracy.
The observation log is given in Table 2.
Exposure times are  $4 \times 210$ s for $g$ filter,
and $4 \times 600$ s for $r$ filter.
Each image was taken with dithering of $20 \sim 30$ arcsec.
These dithered observations turned out to be very useful to
fill the gaps in the mosaic CCDs in the same row. The dithering pattern
used in the present observation, however, could not cover the 
large gaps ($\sim 0.9$ arcmin) between different rows of CCDs,
a gap between the first and the second rows and a gap between the third
and the fourth rows. This uncovered sky region will reduce the efficiency of
finding satellite galaxies, especially small systems (smaller than $20''$).
The seeing ranged from 0.58 to 1.16 arcsec with a median value of 0.74 arcsec.

\subsection{Photometry}

The raw images were preprocessed using Elixir system by CFHT staff. 
Elixir is a collection of programs, databases, and other tools specialized 
in processing and evaluation of the large mosaic data.
Detailed information about the Elixir system can be found in \citet{mag04}.

We combined the dithered exposures of each CCD to get rid of bad pixels,
defective columns and spurious cosmic rays by taking a median of pixel 
values. 
We utilize the resulting images for the detection of point source objects with 
the digital photometry program DAOPHOT \citep{ste94}.
We adopt 3$\sigma$ as a detection threshold.
Instrumental magnitudes of the detected objects 
in the images are derived using the aperture photometry with 
aperture radii of 0.6 to 1.2 arcsec depending on the
seeing FWHM of a point source on each CCD.
The value of the aperture correction was derived from 
the difference between these small aperture
magnitude and large aperture magnitude for well isolated bright stars.

\begin{figure*}
 \epsfig{figure=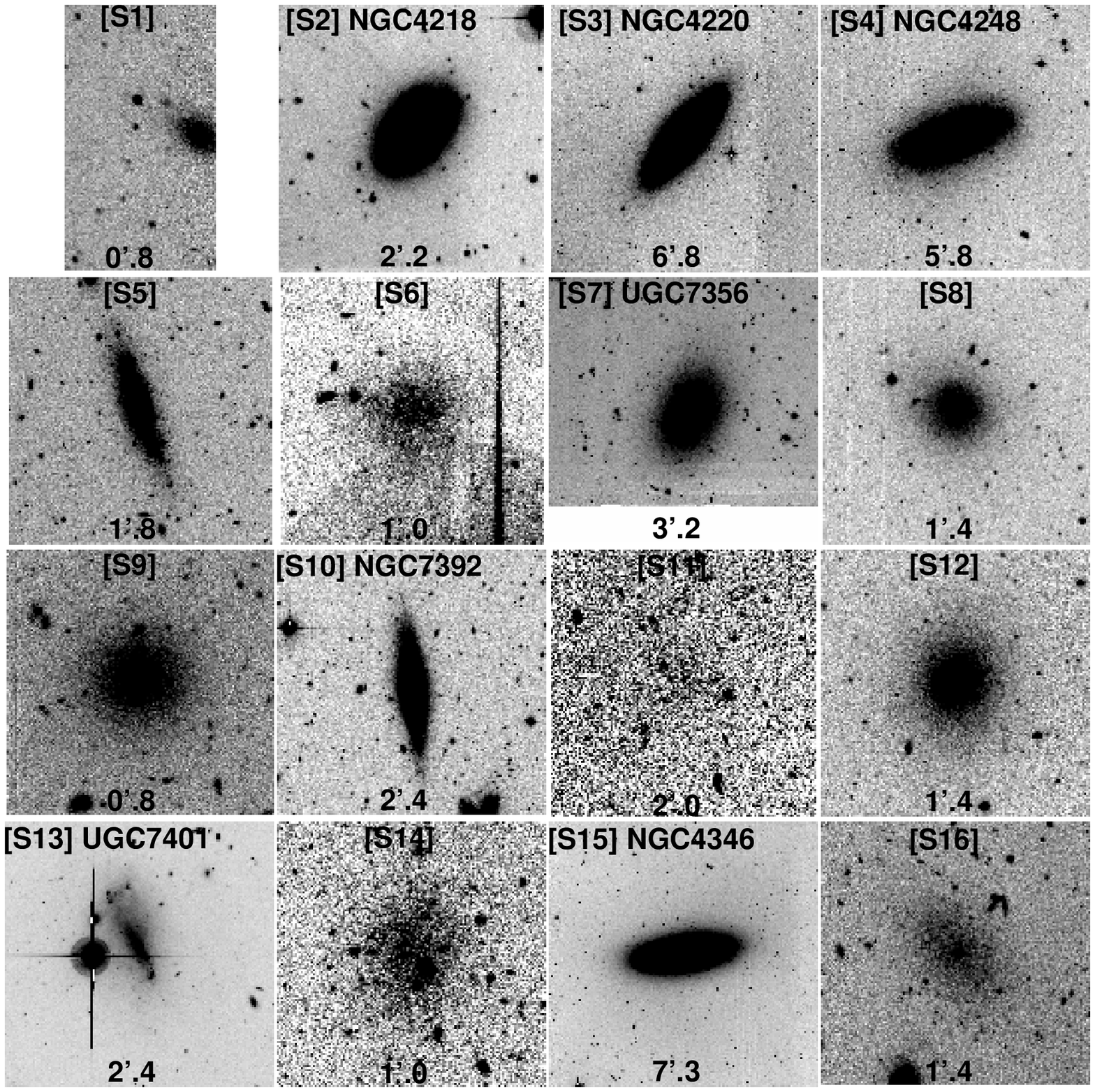, height=0.990\textwidth, width=0.990\textwidth}
 \vspace{3mm}
 \caption{
  An atlas of the satellite galaxies in M106.
  The numbers at the bottom of individual panels represent the size of the field of view.}
 \label{fig_atlas}
\end{figure*}

To derive the radial surface brightness profiles of the satellite galaxies 
we perform surface photometry.
We define the rough boundary of the satellite galaxies by estimating major axis radius, 
ellipticity and
position angle for each satellite galaxy (4th, 5th and 6th columns in Table 3) 
by visual inspection.
Then we obtain the radial surface brightness profiles of the satellite 
galaxies using a custom software
{\bf{\it Ellipsephot}}. {\bf{\it Ellipsephot}} performs a radial 
surface photometry using elliptical apertures by dividing
an elliptical annulus into sectors to deal with unexpected spurious 
sources and/or bad columns. For the surface photometry of candidate
satellite galaxies of M106 we divide an elliptical annulus into 8
sectors and take the median value of surface brightness of these sectors as the
surface brightness of certain radius. This process is useful
to remove the effect of spurious sources and/or bad columns in the surface
photometry of a galaxy. Preliminary version of
{\bf{\it Ellipsephot}} was successfully applied in a 
study of the surface photometry of dwarf galaxies
in the Local Group \citep{lee99}. {\bf{\it Ellipsephot}} is
useful to obtain the radial surface photometry
of non-elliptical shaped and/or very low surface brightness galaxies,
because it is not involved with fitting process. Therefore, {\it Ellipsephot}
can be used for the surface photometry of all extended sources 
irrespective of their shapes.
Users need to determine the centre, ellipticity, and
position angle of the extended structure to run {\it Ellipsephot}.
For galaxies with elliptical structures
we obtain their surface photometry using IRAF/STSDAS/ellipse task. 
IRAF/ellipse task performs
ellipse fitting to the observed isophotes \citep{jed87}. 
A comparison of the surface photometry
for elliptical galaxy NGC 4346 (S15, see Table 3) shows an 
excellent agreement between the 
results using {\it Ellipsephot} and
that of IRAF/ellipse. 

\subsection{Standard Calibration}

CFHT staff provided the transformation equations 
for a standard calibration from the photometry of the 
standard stars observed during the same night.
We obtained the instrumental magnitudes of the standard stars 
using the aperture radius of 7.5 arcsec as used in \citet{gei96}.
The standard transformation equations are: 
$g = g' + 0.148 (g'-r') -0.150 X + const $ 
with rms=0.019 and N=72, and
$r = r' + 0.000 (g'-r') -0.100 X + const $  with rms=0.008 and N=66, 
where the lower case letters represent the standard magnitudes, 
the primed lowercase letters the instrumental magnitudes 
(with a DAOPHOT system zero point of 25.0), and $X$ the air mass.
Zero-points of these transformation relations are slightly different for
different observed field.
We transformed the instrumental magnitudes of the sources detected in
fields F1 and F2 onto the standard
system using these transformation equations. The instrumental magnitudes of the point sources detected in F3 and F4 were transformed onto 
the standard system by comparing
the magnitudes of the point sources common between F1 and F3 and 
between F2 and F4, respectively.

SDSS Data Release 5 (DR5, \citet{ade07}) covers the entire field of the present 
MegaCam observation of M106, though SDSS observation is significantly
shallower than MegaCam observation. We find common point sources 
between SDSS DR5 catalogue and this study, and
derive the magnitude difference between the transformed standard magnitudes of this study and
the point-spread function fitting magnitudes of point sources in SDSS DR5.
We adjust the standard magnitudes of the point sources by adding 
these magnitude differences
in each CCD image.

\begin{figure*}
 \epsfig{figure=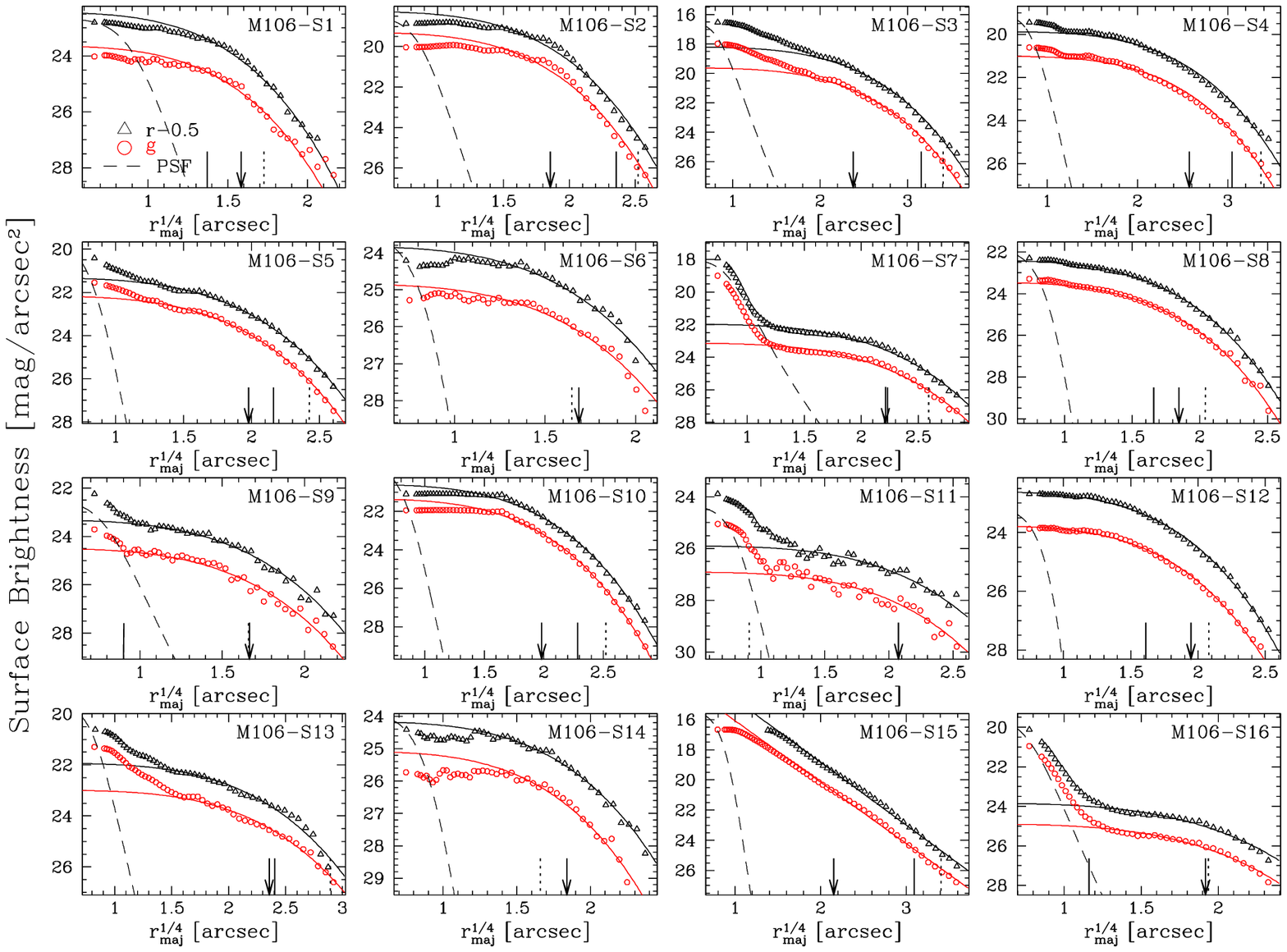, height=0.830\textwidth, width=0.990\textwidth}
 \vspace{3mm}
 \caption{
  Surface brightness profiles of the candidate satellite galaxies in 
M106 against the semi-major axis radii.
  The curved broken lines represent the point spread function.
  Locations of the effective radii, standard radii and Holmberg radii
  are marked with vertical arrows, vertical solid lines, and vertical dotted lines, respectively.
  The $r$-band surface magnitude is shifted by -0.5 mag for easier comparison.
  Surface brightness profiles are fitted by the exponential law except for S15,
  whose profiles are
  fitted with de Vacouleurs law.
  The curved solid lines over the data points represent the best-fitting with the
  exponential law or de Vacouleurs law.}
 \label{fig_surf}
\end{figure*}

\begin{figure*}
 \epsfig{figure=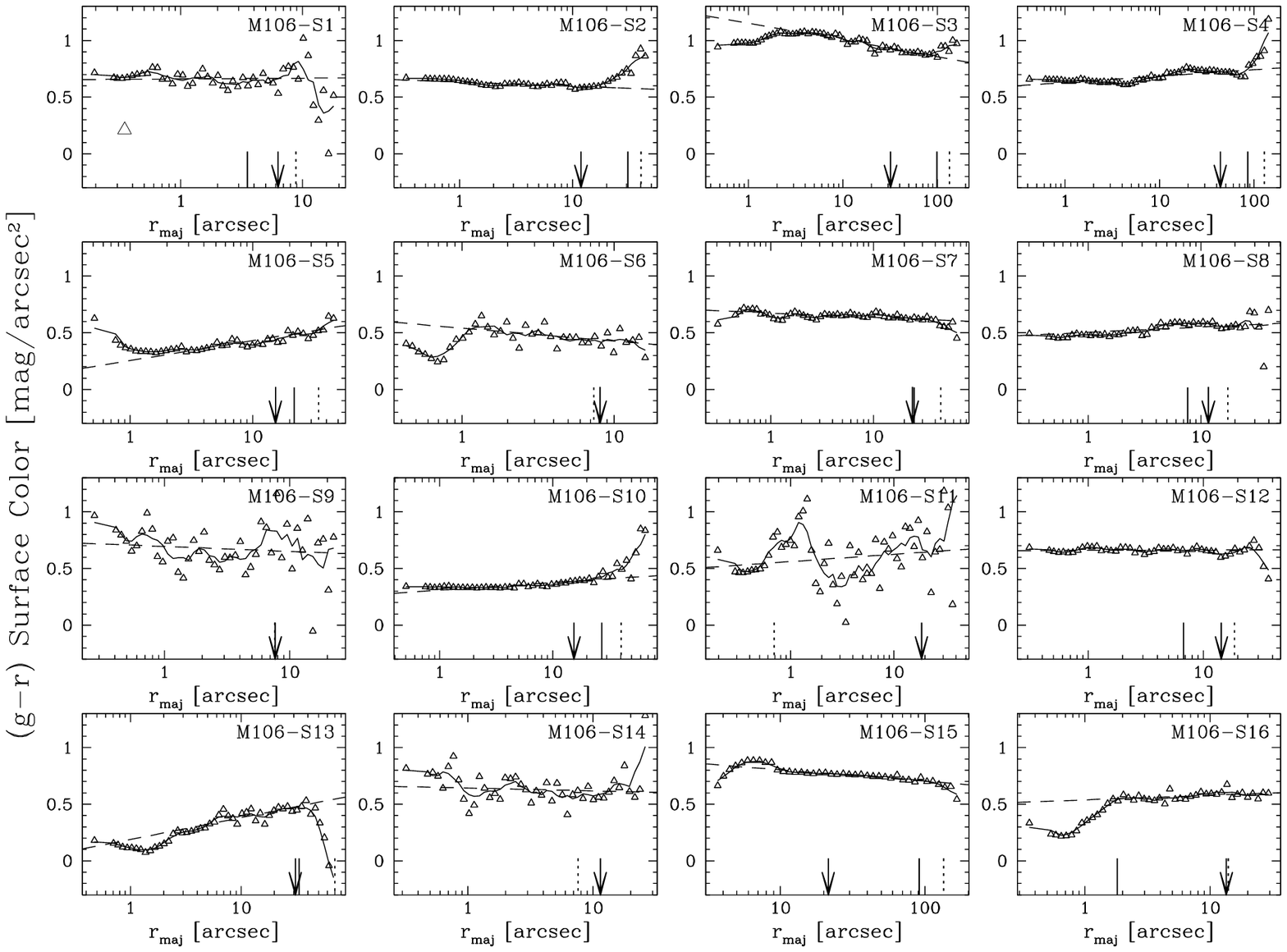, height=0.830\textwidth, width=0.990\textwidth}
 \vspace{3mm}
 \caption{
  $(g-r)$ surface colour profiles of the satellite galaxies in M106. Solid lines
  represent the box-car smoothed profiles of the observed data. Dashed lines
  represent the linear fits of the surface colour profiles for the intermediate
  radial range (see Table 4 for details).
  in each panel. The typical error of the surface colour ranges from 0.1 to 0.2.
  The meaning of the vertical arrows and lines are the same as those in Fig. 3.}
 \label{fig_scolor}
\end{figure*}

\subsection{Sky Brightness}

Sky brightness is a critical parameter to find faint 
satellite galaxy. Detection efficiency of
the extended sources (e.g., galaxies) strongly depends on both surface 
brightnesses of the sources
and the background, while that of the of point sources
depends mainly on the total source brightness. Therefore it is important 
to carry out observations on
the dark night to find as many faint satellite galaxies as possible.

We measure the sky brightnesses of the individual fields,
and list their values in Table 2.
We select a few sky regions around the satellite galaxies to measure
the sky brightness located $\sim3-5$ times of the outer radii
(see Table 3 for the values of the outer radii of the satellite galaxies) where
no foreground and background sources are included.
The mean sky brightnesses of F1 and F2 in $g$ and 
$r$ bands are 22.31 and 21.54 mag arcsec$^{-2}$,
respectively, corresponding to 22.78 and 21.86 mag arcsec$^{-2}$ 
in $B$ and $V$ bands. We transform the
SDSS $g$ and $r$
magnitudes to Johnson $B$ and $V$ magnitudes using the 
transformation relation in \citet{cho08},
derived by comparing SDSS DR5 magnitude of standard stars of \citet{lan92}
to their standard magnitude.
The derived sky brightness is slightly fainter than the 
typical dark sky brightness at the
CFHT site (22.00 and 21.40 mag arcsec$^{-2}$ in $g$ and $r$ bands).
The mean sky brightness of F3 is
significantly brighter than those of F1 and F2 by 0.8 and 0.6 mag 
in $g$ and $r$ bands, respectively.
Sky brightness of F4 is slightly fainter than that of F3,
but is brighter than those of F1 and F2 by $\sim0.5$ mag.
The cause of the relatively brighter mean sky brightness of 
F3 compared to F1 and F2 might be due to the scattered moon light by
the cloud and rather smaller lunar angle, and larger lunar phase.
The extinction by cloud for F3 might affect the detection limiting
magnitude and limiting surface brightness, therefore, the detection
efficiency of satellite galaxies in F3 might be lower than those
of the other fields.
We list the lunar angle and the
lunar phase of each observation field in Table 2.

\begin{table*}
 \centering
 \begin{minipage}{180mm}
 \raggedright
  \caption{A list of the candidate satellite galaxies in M106}
  \begin{tabular}{@{}cccccccccccc@{}} \hline\hline
   ID & R.A. & Dec. & $R_{\rm out}$ & $\epsilon_{\rm out}$ & $PA_{\rm out}$ &
   $g^T$ & $r^T$ & ${\it{V}}_{\rm r}$ & Name & Type & Selection$^a$ \\ \hline\hline
    S1 & 12 14 05.00 & 47 26 08.6 &  12.9 & 0.329 & 146 & 19.20$\pm$0.06 & 18.47$\pm$0.05 &    --     & --      & dE        & 2 \\
    S2 & 12 15 46.41 & 48 07 51.0 &  39.8 & 0.294 &  48 & 13.49$\pm$0.06 & 12.88$\pm$0.05 &   766$^b$ & NGC4218 & Sd        & 1,3 \\
    S3 & 12 16 11.71 & 47 52 59.7 & 124.5 & 0.636 &  50 & 11.88$\pm$0.04 & 10.95$\pm$0.03 &   920$^b$ & NGC4220 & Sa        & 3 \\
    S4 & 12 17 49.90 & 47 24 33.0 & 113.0 & 0.553 &  20 & 12.86$\pm$0.24 & 12.12$\pm$0.16 &   484$^c$ & NGC4248 & Im        & 1,3 \\
    S5 & 12 18 11.07 & 46 55 00.7 &  41.0 & 0.725 & 110 & 16.29$\pm$0.06 & 16.29$\pm$0.09 &   431$^b$ & --      & Sm        & 1,3 \\
    S6 & 12 19 06.21 & 47 43 49.8 &  10.3 & 0.100 & 170 & 19.30$\pm$0.06 & 18.74$\pm$0.08 &    --     & KK132   & dSph      & 1 \\
    S7 & 12 19 09.39 & 47 05 27.5 &  48.1 & 0.250 &  70 & 15.37$\pm$0.03 & 14.75$\pm$0.03 &   202$^b$ & UGC7356 & dE,N      & 1,3 \\
    S8 & 12 19 33.21 & 47 27 05.3 &  16.7 & 0.111 & 170 & 17.32$\pm$0.22 & 16.76$\pm$0.44 &   780$^d$ & KK134   & dE        & 2 \\
    S9 & 12 19 35.96 & 47 05 35.2 &  11.1 & 0.000 &   0 & 18.82$\pm$0.10 & 18.13$\pm$0.05 &    --     & --      & dSph      & 2 \\
   S10 & 12 20 17.49 & 48 08 15.6 &  55.6 & 0.727 &  97 & 15.95$\pm$0.04 & 15.56$\pm$0.03 &   797$^c$ & UGC7392 & Sd        & 1,3 \\
   S11 & 12 20 30.07 & 47 29 24.8 &  18.5 & 0.450 & 140 & 20.10$\pm$0.54 & 19.42$\pm$0.45 &    --     & --      & dSph/dIrr & 1 \\
   S12 & 12 20 40.18 & 47 00 03.2 &  22.5 & 0.231 &  80 & 17.41$\pm$0.03 & 16.69$\pm$0.02 &    --     & KK136   & dE        & 2 \\
   S13 & 12 20 48.42 & 47 49 33.4 &  39.3 & 0.667 & 115 & 15.66$\pm$0.20 & 15.23$\pm$0.21 &   762$^b$ & UGC7401 & Im        & 1,3 \\
   S14 & 12 20 54.86 & 46 49 48.8 &  13.1 & 0.214 &  45 & 19.22$\pm$0.16 & 18.49$\pm$0.16 &    --     & --      & dSph      & 2 \\
   S15 & 12 23 27.94 & 46 59 37.8 & 121.1 & 0.539 & 187 & 11.69$\pm$0.04 & 10.96$\pm$0.04 &   770$^c$ & NGC4346 & S0/SB0    & 1,3 \\
   S16 & 12 23 46.18 & 47 39 31.9 &  22.3 & 0.417 & 140 & 18.39$\pm$0.09 & 17.82$\pm$0.09 &    --     & --      & dE,N      & 1 \\
   P1  & 12 18 18.76 & 47 48 17.0 &   6.5 & 0.000 &   0 & 19.86$\pm$0.16 & 19.86$\pm$0.12 &    --     & --      & dsph      & - \\
   P2  & 12 18 06.54 & 46 28 50.1 &   2.7 & 0.188 & 135 & 20.96$\pm$0.17 & 21.07$\pm$0.22 &    --     & --      & dIrr/dSph & - \\
   P3  & 12 17 06.33 & 46 54 08.2 &   2.2 & 0.146 & 130 & 19.94$\pm$0.07 & 20.19$\pm$0.07 &    --     & --      & dIrr/dSph & - \\
   P4  & 12 17 02.25 & 46 54 20.2 &   4.1 & 0.475 & 172 & 19.77$\pm$0.06 & 19.87$\pm$0.06 &    --     & --      & dIrr/dSph & - \\
   P5  & 12 16 37.40 & 47 09 12.0 &   1.8 & 0.000 &   0 & 20.33$\pm$0.10 & 20.69$\pm$0.23 &    --     & --      & dIrr/dSph & - \\
   \hline
  \end{tabular}

  $^a$Applied criteria for individual candidate satellite galaxies.

  $^b$An average value of NED and SDSS.

  $^c$Radial velocity from NED.

  $^d$Radial velocity from SDSS.
 \end{minipage}
\end{table*}

\section{Results}

\subsection{Selection of Satellite Galaxies}

Distinguishing the satellite galaxies from the background galaxies and foreground objects is
one of essentials in the present study. Based on the simple hypothesis of similarity in physical
properties of the satellite galaxies of M106 and the Local Group,
the satellite galaxies 
of M106 are expected to have a variety of size, total magnitudes, 
colours and surface brightnesses.
We therefore, need to explore a large set of searching parameters to
complete the finding of satellite galaxies of M106. Most efficient and complete method
suitable for finding satellite galaxies in the wide field survey is
human visual inspection. 
Three of the authors search both the combined fits images and true color images
independently and repeat the searching process until three authors agree one
another. The galaxies identified as satellite galaxies by either one or two authors
are classified as probable satellite galaxies.

All the extended sources detected in the images might be potential satellite galaxy
candidates of M106. However, considering the total mass 
(similar to or smaller than the mass of Milky Way), 
M106 is expected to host a few tens of satellite galaxies. Therefore many of the
extended sources in the images are not the satellite galaxies of M106. To 
select genuine satellite galaxies while keeping as many candidate galaxies
as possible,
we construct the following steps to find the satellite galaxies.

\noindent
(1) If a galaxy that shows some resolved stars, it is considered to
be a satellite galaxy. Since we can resolve a significant portion 
of M106 into individual sources,
the partially resolved galaxies near M106 could be the most plausible 
candidates of satellite galaxy of M106.

\noindent
(2) We consider galaxies with no resolved structure but showing rather extended ($\ge 10''$)
and faint surface brightness structure (extremely faint surface brightness) as 
candidate satellite galaxies.

\noindent
(3) Galaxies with similar radial velocity to that of M106 are considered to be 
satellites.

Finally we find 16 satellite candidates and five probable satellite candidates of M106.
Hereafter, we call the candidate satellite galaxies as satellite galaxies.
Twelve of the satellite galaxies are  previously known, and four of
them are new findings.
We list the properties of these satellite galaxies in Table 3.
The galaxy type is estimated based on the morphology of a galaxy on CCD images 
and the surface brightness profiles. 
We display an atlas of 16 satellite galaxies in Fig. 2. Relatively large
satellite galaxies (S2, S3, S4, S7, and S15) show elliptical shapes. 
Two intermediate-size galaxies,
S5 and S10 are typical edge-on disk galaxies. 
S13 has a structure of typical irregular galaxy, being close to a very
bright foreground star. We masked out the sky regions of this bright 
star before deriving the surface photometry.
Two satellite galaxies S6 and S9 showing spherical shapes with relatively low central
surface brightness seem to be dwarf spheroidal galaxies.
Surface brightness profiles of satellite galaxies S8 and S9 with 
spherical appearance are
well represented by exponential profiles for the entire radial ranges.
Two satellite galaxies showing spherical shapes with extremely low 
surface brightness,
S11 and S14, are morphologically similar to
dwarf spheroidal galaxies. S16 shows an elongated shape and its surface 
brightness profiles for the outer region
are well represented by an exponential law. 
S16 is a dwarf elliptical galaxy with a central nucleus.
S1 was located at the boundary of
CCD chip, so that we could not estimate its morphological type.
The surface photometry of S1 indicates that
the morphological type of S1 is close to a dwarf elliptical galaxy.
We classify P1 as
probable satellite galaxy of M106 because its central surface brightness is 
much fainter than
all the satellite galaxies. Remaining galaxies, P2, P3, P4, and P5, which have
extremely small effective radii ($<3.5$ arcsec), are also classified as probable
satellite galaxies of M106.

\begin{table*}
 \centering
 \begin{minipage}{127mm}
 \raggedright
  \caption{Surface Photometry of satellite galaxies in M106}
  \begin{tabular}{@{}ccccccccc@{}} \hline\hline
   ID & $\Sigma_{0,V}^a$ & $R_{\rm eff}^b$ & $R_{25}^b$ & $R_{\rm holm}^b$ &
   $H_g^c$ & $H_r^c$ & Fit ranges$^d$ & $(g-r)$ slope$^e$ \\ \hline\hline
   S1  & 23.69$^4$ &  6.3 &   3.5 &   8.9 &  4.09$\pm$0.17 &  3.99$\pm$0.16 & 0.2$\sim$10  &  0.018$\pm$0.119 \\
   S2  & 19.63$^4$ & 11.9 &  30.8 &  40.6 &  6.82$\pm$0.10 &  7.11$\pm$0.11 & 0.2$\sim$20  & --0.035$\pm$0.078 \\
   S3  & 17.46$^3$ & 32.0 &  99.3 & 134.6 & 22.88$\pm$0.40 & 22.89$\pm$0.40 & 2.0$\sim$100 & --0.147$\pm$0.066 \\
   S4  & 20.24$^4$ & 43.9 &  86.0 & 128.7 & 27.02$\pm$0.30 & 28.55$\pm$0.33 & 0.3$\sim$90  &  0.057$\pm$0.040 \\
   S5  & 21.43$^3$ & 15.3 &  21.8 &  34.5 &  9.54$\pm$0.21 & 10.05$\pm$0.24 & 1.5$\sim$50  &  0.176$\pm$0.090 \\
   S6  & 25.00$^4$ &  8.1 &    -- &   7.4 &  6.85$\pm$0.30 &  6.42$\pm$0.27 & 1.0$\sim$12  & --0.115$\pm$0.172 \\
   S7  & 20.71$^2$ & 23.9 &  24.8 &  44.9 & 16.76$\pm$0.46 & 16.30$\pm$0.44 & 0.2$\sim$50  & --0.041$\pm$0.046 \\
   S8  & 23.18$^1$ & 11.6 &   7.6 &  17.3 &  7.25$\pm$0.10 &  7.27$\pm$0.10 & 0.2$\sim$40  &  0.045$\pm$0.053 \\
   S9  & 23.79$^2$ &  7.7 &   0.7 &   7.6 &  5.94$\pm$0.14 &  5.90$\pm$0.14 & 0.2$\sim$25  & --0.028$\pm$0.071 \\
   S10 & 21.74$^4$ & 15.5 &  27.3 &  40.4 &  9.28$\pm$0.19 &  9.88$\pm$0.22 & 0.3$\sim$30  &  0.069$\pm$0.075 \\
   S11 & 25.59$^3$ & 18.5 &    -- &   0.7 & 16.43$\pm$0.63 & 18.40$\pm$0.80 & 0.2$\sim$30  &  0.054$\pm$0.061 \\
   S12 & 23.50$^1$ & 14.4 &   6.8 &  18.7 &  9.19$\pm$0.14 &  9.11$\pm$0.14 & 0.2$\sim$30  &  0.003$\pm$0.061 \\
   S13 & 21.25$^3$ & 30.9 &  33.5 &  70.2 & 22.46$\pm$0.87 & 20.34$\pm$0.69 & 2.5$\sim$40  &  0.193$\pm$0.140 \\
   S14 & 25.44$^4$ & 11.4 &    -- &   7.6 &  7.49$\pm$0.26 &  8.91$\pm$0.36 & 0.3$\sim$25  & --0.016$\pm$0.067 \\
   S15 & 16.70 & 21.5 &  91.6 & 135.3 &      --        &     --        & 10.0$\sim$150 & --0.060$\pm$0.151 \\
   S16 & 22.12$^2$ & 13.5 &   1.8 &  14.0 & 11.91$\pm$0.61 & 12.35$\pm$0.65 & 2.0$\sim$30  &  0.041$\pm$0.139 \\
   P1  & 25.77 &  9.2 &    -- &   1.3 &  5.39$\pm$0.34 &  3.87$\pm$0.17 & 0.5$\sim$20  &  -0.328$\pm$0.074 \\
   P2  & 23.80 &  1.8 &   --  &   4.9 &  1.03$\pm$0.04 &  0.73$\pm$0.02 & 0.5$\sim$4   &  -0.489$\pm$0.239 \\
   P3  & 22.99 &  2.3 &   3.2 &   5.1 &  1.71$\pm$0.04 &  1.60$\pm$0.03 & 0.5$\sim$10  &  -0.213$\pm$0.111 \\
   P4  & 22.80 &  2.5 &   4.2 &   5.8 &  1.60$\pm$0.03 &  1.60$\pm$0.03 & 0.5$\sim$10  &  -0.156$\pm$0.111 \\
   P5  & 24.59 &  3.4 &   1.3 &   5.3 &  2.90$\pm$0.24 &  2.43$\pm$0.17 & 0.5$\sim$6   &  -0.301$\pm$0.170 \\
   \hline
  \end{tabular}

  $^a\Sigma_{0,V}$ denotes the central surface brightness in units of mag arcsec$^{-2}$.

  $^b$Effective radius($R_{\rm eff}$), standard radius($R_{25}$), and Holmberg radius($R_{\rm holm}$)
  measured in $r$-band images are given in units of arcsec.

  $^c$Disk scale length($H$) in $g$ and $r$ bands in units of arcsec.

  $^d$Radial ranges for linear fit of ($g-r$) colour profiles in units of arcsec.

  $^e$($g-r$) colour gradient in units of mag/$\Delta$logr where r in units of arcsec.

  $^{1,2,3,4}$ Categories for surface brightness profiles and central surface brightness 
   as described in \S3.2.
 \end{minipage}
\end{table*}

\subsection{Surface Brightness and Color Profiles}

We display the radial surface brightness profiles of all the satellite galaxies of 
M106 in Fig. 3, and list the results of surface photometry including $(g-r)$ colour 
gradient in Table 4.
The radial profiles of surface magnitude in $g$ and $r$ bands are displayed in
open circles and open squares, respectively.
We marked the position of the effective radius ($r_{\rm eff}$) within which 
half of the total galaxy luminosity is emitted by thin arrows, together with
the positions of two characteristic radii, $r_{25}$ and $r_{\rm Holm}$,
by the vertical solid lines and the vertical dotted lines.
$r_{25}$ and $r_{\rm Holm}$ are defined as the radii where the $B$-band surface
brightness becomes $25$ mag arcsec$^{-2}$ and $26.5$ mag arcsec$^{-2}$,
respectively.

Most of the satellite galaxies of M106 show surface brightness profiles 
typical of a disk galaxy, while the surface brightness profiles of S15
resemble those of an elliptical galaxy.
Surface brightnesses of three satellite galaxies S6, S11, and S14 are extremely 
low so that we could not determine the two characteristic 
radii, $r_{25}$, and $r_{Holm}$
for these galaxies. These low surface brightness galaxies are expected to be an analogy of 
Sextans and Sagittarius dwarf spheroidal galaxies in the Local Group.
Radial surface brightness profiles of the satellite galaxies, S7, S9, and S16 show a strong
signature of the presence of the central nucleus.
The radial surface brightness profile of S7 near the galaxy centre 
is found to be very close to that of the point source profile,
implying the presence of the central nucleus.
On the other hand, the central structures of S9 and S16 are slightly extended compared
with the typical point source profile.

We fit the derived radial surface brightness profiles with a single exponential law
(thin solid lines over data points) except for
S15, for which we use the de Vaucouleures law. 
The physical properties of the 
satellite galaxies of M106 derived by fitting the radial surface brightness 
profiles are listed in Table 4. 
The radial surface brightness profiles of the outer regions of the
satellite galaxies of M106 are fitted well by an exponential profile.
However, near the galaxy
centre there are seen a variety of structures. We divide 
the satellite galaxies of M106 with 
exponential disk into 4 categories based on their 
central surface brightness profiles: 
(1) galaxies with surface brightness profiles represented 
well by a single exponential law (S8 and S12), 
(2) galaxies expected to host a clear/marginal central nucleus (S7, S9, and S16), 
(3) satellite galaxies showing brighter central surface brightness than the exponential law but 
more extended than the PSF (S3, S4, S5, S11, and S13), and 
(4) galaxies with fainter central brightness than the exponential disk profiles
(S1, S2, S6, S10 and S14).

The central structures of the satellite galaxies in category 3 might be explained by the presence of
the central bulge in these galaxies. The appearances of these galaxies are, however, quite different from
one another. Moreover, two galaxies, S5 and S10, showing typical shapes of disk edge have quite different
central structure. 

We present the derived radial $(g-r)$ colour profiles of the satellite galaxies of M106 in Fig. 4.
For clear appearance we apply the box car smoothing kernel 
to the data (solid lines). $(g-r)$ surface colour of the satellite galaxies
ranges  from 0.1 to 1.0 mag arcsec$^{-2}$, corresponding to 
0.2 and 1.2 mag arcsec$^{-2}$ in $(B-V)$. We note that
the expected foreground reddening toward M106 is negligible ($E(B-V)<0.02$, \citet{sch98}). 
We plot the radial colour profiles of 16
satellite galaxies using the same scale in y-axis for clear comparison. We derive the colour 
gradient of $(g-r)$
colour profiles of these galaxies and summarized the results in Table 4. For the linear fit 
of the surface colour profile we only include
the data in the intermediate radial range as summarized in Table 4. The surface colour 
profiles of S3 and S15 get bluer with
increasing radius, while those of S5 and S13 get redder as radius increases. The radial 
surface colour profiles of remaining
12 galaxies do not show any clear signature of colour gradient. Despite the similar trend in 
the radial surface brightness profiles,
the colour profiles of S3 and S5 differ clearly: slow increase in $(g-r)$ colour with 
increasing radius near
the central region followed by a slower decrease (getting bluer) for S3, 
but the trend is quite opposite for S5.
The marginally visible resolved stars in S5 might indicate the presence 
of young MS stars and/or AGB stars, 
which are brighter than the tip of the red giant branch stars. This
indicates the recent star formation in S5.
For three galaxies (S6, S11, and S14) with extremely low surface brightness,
we found no trend in the colour profiles due to the large photometric errors.

\subsection{Spatial Distribution}

The two dimensional distribution of the 16 satellite galaxies of M106 (see Fig. 1) shows (1) the central concentration of the satellite galaxies
toward M106 and (2) the higher probability to find satellite  galaxies
along the major axis of M106.
We investigate these features in more detail using quantitative analysis as 
shown in Fig. 5.

\begin{figure}
 \epsfig{figure=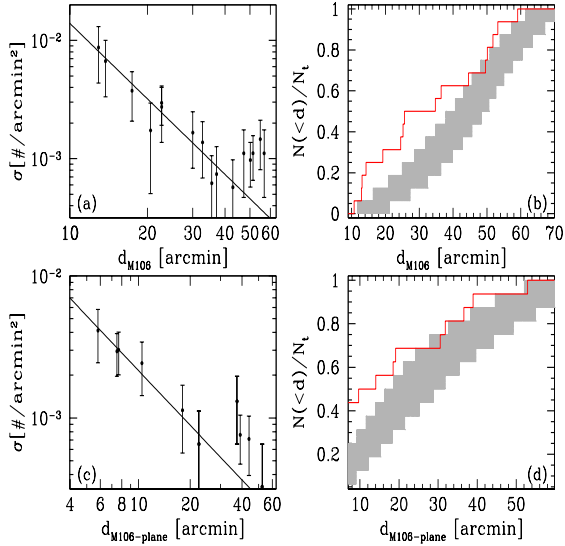, height=0.480\textwidth, width=0.480\textwidth}
 \vspace{3mm}
 \caption{
  Spatial distributions of the satellite galaxies of M106: 
  (a) number density distribution against the projected distance from the
  centre of M106, (b) cumulative number distribution of the
  galacto-centric radial distance of the satellite galaxies from M106,
  (c) number density distribution of the satellite galaxies against the projected
  distance from the major axis of M106, and (d) cumulative number distribution
  with increasing projected distance from the major axis of M106.
  The solid lines in Figs. 5(a) and 5(c) are the best-fitting power-law
  for the ranges $0<d_{M106}<45\prime$, and $0<d_{M106-plane}<30\prime$.
  The shaded regions in Figs. 5(b) and 5(d) represent
  the expected cumulative number distribution with 1$\sigma$ boundary assuming a
  spatially uniform distribution.}
 \label{fig_gdist}
\end{figure}

We show one dimensional spatial distributions of the 16 satellite galaxies in Fig. 5.
Fig. 5 (a) and (c) present the surface number density profiles as a function of the projected
distance from the centre of M106 and from the major axis of M106.
Both profiles
show clear signatures of concentration toward the centre of M106 and 
toward the major axis of M106, respectively. We fit the radial
number density profile of the satellite galaxies using a power-law for the 
radial range of $0\sim45$ arcmin, finding a power index of $-2.12\pm0.55$. 
A power index
of a power-law profile for Fig. 5(c) is $-1.28\pm 0.52$ 
much flatter compared to the power index for galacto-centric
radial distribution (Fig. 5(a)). The derived power-law index for the 
galacto-centric surface number density profile is very similar to the
expected value for isothermal distribution. This implies that 12 of
16 candidate satellite galaxies with $d_{M106}<50$ arcmin
($\approx100$ kpc) 
found in this study are probably genuine satellite galaxies of M106.

Because we are dealing with small number statistics ($N<20$), the derived 
power index
of power-law for the number density profiles might be affected significantly by
the binning method. To examine the effect of binning for the number 
density profiles we
derive the number density profiles using four different binning methods; (1) an equal number of data
points in each bin, (2) binning the data points in an equal linear 
interval, (3) binning the
data points in an equal logarithmic interval, and (4) binning the data 
points in an equal $r^{1/4}$ interval.
We applied four different binning methods to obtain 
the radial number density profiles of 16 satellite galaxies of M106.
We then, fit the data with $d_{M106} < 45$ arcmin to the power-law for the each
density profile, obtaining power indices of $-2.90\pm0.89$, 
$-2.14\pm0.74$, $-1.93\pm0.51$, and $-2.21\pm0.59$, respectively. 
All of the power indices agree well within the errors, and the derived
power-index ($-2.12$) is close to the weighted mean of the values.

The satellite galaxies in the outer region ($>100$ kpc) show more or less like 
a flat distribution rather than the power-law. 
We fit the radial number density profile of the satellites including
the galaxies in the outer region of M106 ($d_{M106}>100$ kpc),
obtaining a power index of $-1.33\pm 0.36$.
This is somewhat shallower than the power index for the
radial number density profile of 12 satellite galaxies.
However due to the large uncertainty
it is quite uncertain whether the satellite distribution 
of the outer region of M106 is flat or not. 
One of the causes for the flat distribution of
M106 satellite located at the outer region may be the 
inclusion of the background galaxies in the list of the
M106 satellite galaxies.

 Fig. 5 (b) and (d) show the cumulative number distribution of 
satellite galaxies against the projected distance from the centre
of M106, and 
that of the projected distance from the major axis of M106, respectively 
(solid lines). For comparison we display the expected
cumulative number distribution of 16 artificial sources assuming 
spatially uniform distribution with shaded regions. The width of the shaded regions
represents $1\sigma$ boundary. The concentrated distribution of the 
satellite galaxies toward the centre of M106 is also shown
clearly with a high statistical confidence ($>95\%$). For the outer 
region ($d_{M106}>50$ arcmin) the observed distribution of
the satellite galaxies is very similar to that of the uniform 
distribution (This is also observed in Fig. 5 (a)). In addition to
the central concentration toward the centre of M106, they are 
located preferentially along the major
axis of M106. This implies that a large portion of the 
satellite galaxies of M106 were formed 
around M106 and are captured by M106
moving along the merging filaments, which are spatially 
aligned with the major axis of M106.

\begin{figure}
 \epsfig{figure=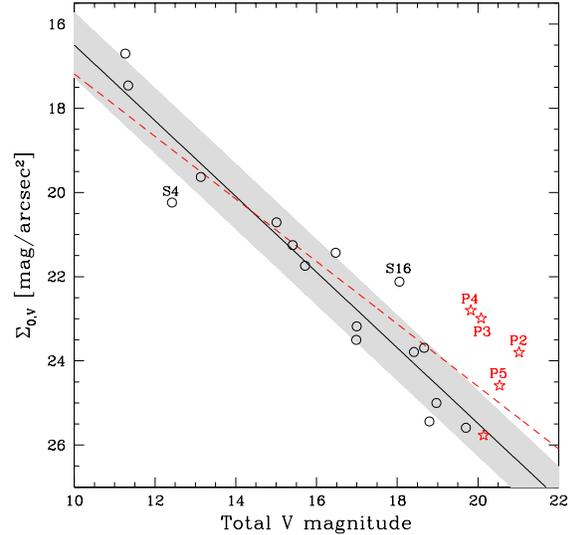, height=0.480\textwidth, width=0.480\textwidth}
 \vspace{3mm}
 \caption{
  Distribution of both the candidate satellite galaxies (open circles)
  and the probable satellite galaxies (star symbols) of M106 in
  total magnitude versus central surface brightness ($\Sigma_{0}$) in 
 $V$-band.
  The solid
  and dashed lines represent the best-fit linear relations
  for the data of 16 candidate satellite galaxies and
  16+5 satellite galaxies, respectively.
  The hashed region around the solid line denotes $1\sigma$ boundary
  of the best-fit linear relation.}
 \label{fig_SBTmag}
\end{figure}
 
\section{Discussion} 

\subsection{Misidentification and Missing Satellite Galaxies}

The sample of 16 satellite galaxies of M106 found in this study might 
contain non members.
Most of the misidentifications are
expected to be connected with faint background galaxies. The probability 
to include background galaxies as the satellite galaxies of M106
can be higher in applying the criterion (3) in \S 3.1 because the 
background galaxies (but still within a few Mpc from M106) with large
peculiar motion could possibly be identified as the satellite galaxies of 
M106. In Table 3, there are 9 galaxies with known radial velocities.
Considering the radial velocity of M106
($v_r=448 $kms$^{-1}$), S3 that shows the largest radial velocity difference
($\Delta v_r = 472$km$^{-1}$) still seems to be within the expected
dynamical boundary of 
M106 ($\vert \Delta v_r(sat-host) \vert$ $\approx400-500$kms$^{-1}$).
Thus we conclude that the 
probability including misidentification in
9 satellite galaxies with radial velocity information is negligible.

\begin{figure}
 \epsfig{figure=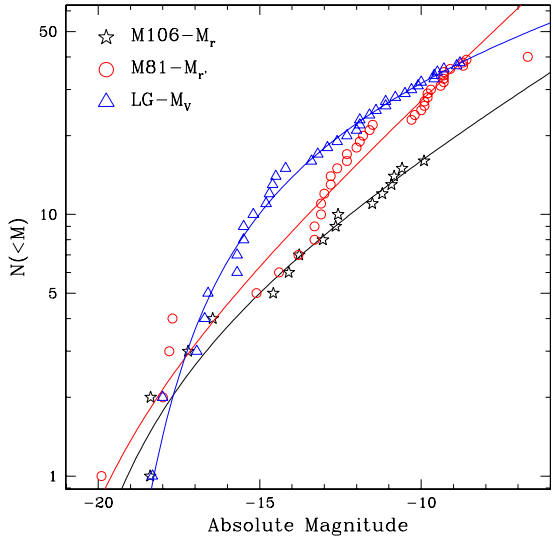, height=0.480\textwidth, width=0.480\textwidth}
 \vspace{3mm}
 \caption{
  Cumulative luminosity function of the satellite galaxies
  in M106 (star symbols), LG (open triangles),
  and the M81 group (open circles). The solid lines overlaid
  on the data-points are the best-fitting cumulative
  Schechter functions. The photometric bands
  for each data-set are noted in the figure.}
 \label{fig_lf}
\end{figure}
 
For the remaining 7 galaxies without radial velocity
information, we are not able to discriminate the genuine satellites 
from the misidentified ones. We need to measusre their radial velocities
by spectroscopic observations or to derive their distances by high resolution 
photometry.
Observations with higher spatial resolution using space 
telescope and/or ground-based observations equipped with
adapted optics at near-infrared wavelengths are crucial to confirm 
these galaxies as the satellite galaxies of M106, by detecting a
significant number of red giant branch stars in the
satellite galaxies.

One of the possible contaminations in the current list of satellite 
galaxies of M106 is the inclusion of background galaxies especially
for the outer region of M106. The number of satellite galaxies at
$d_{M106}>50$ arcmin is four. Two (S2 and S10)
of these 4 galaxies have the radial velocity of $v_r \approx 750$ km/sec, 
somewhat larger than the radial velocity of M106 but
still smaller than or similar to the radial velocity of S3 and S8.
For the remaining two galaxies (S1 and S16) without radial velocity 
information we are not able to decide if they are dynamically bound to M106.

To examine the misidentification of background galaxies as 
the satellite galaxies of M106, we show the central surface brightness
versus total magnitude of both candidate satellite galaxies (open circles)
and probable satellite galaxies (star symbols) in Fig. 6.
Two candidate satellite
galaxies (S4 and S16) are located somewhat outside of the $1\sigma$ boundary, but still within
$\sim2\sigma$ boundary. For the probable satellite galaxies, we find that only P1 is located well
inside the boundary. The fitting error for the best-fit linear relation using all the data 
(16+5 galaxies)
is $\sim40$\% larger than that for the best-fit linear relation using 16 candidate 
satellite galaxies.
S16 shows some resolved stars with an extended halo, which satisfies the 
most strong selection criterion.
S4 (NGC4248) also shows some resolved stars, and the radial velocity 
($484$ km s$^{-1}$) is very
similar to that of M106. Therefore, we conclude that both S4 and S16 are 
candidate satellite galaxies of M106.

In addition to the misidentification of the satellite galaxies of M106, 
there might be a number of galaxies not
included in the current list of satellite galaxies of M106 because of 
incomplete photometry.
We estimate the expected number of missing satellite galaxies due to 
the gaps in the
present CFHT mosaic observations (see \S 2 for the description of gaps). 
The width of the each gap is $\sim 0.9$ arcmin, resulting in
$360$ arcmin$^2$ in the entire observed field.
This corresponds to $\sim3\%$ of the whole observational field of view. 
Since the gap is rather narrow the probability of missing
a large (size $>1-2$ arcmin) satellite galaxy is considerably small.
We estimate the expected number of missing small satellite 
galaxies ($r \le 1$ arcmin) 
located in these gaps as follows:
First we assume a uniform distribution of the satellite galaxies. We then assume 
that if a half of
a galaxy is located in this gap we fail to find this galaxy as the satellite 
galaxy of M106. 
The expected number of missing satellite galaxies due to the gaps of mosaic 
observation is $\sim 0.7$. Because the distribution
of the satellite galaxies of M106 is concentrated toward the centre of 
M106 and the distribution of the gaps is roughly uniform,
the number might be smaller than 0.7. We compared the catalogue of 
dwarf galaxies around M106 of 
Karachentsev et al. (2007) with our list of satellite galaxies. 
We found that all dwarf satellite galaxies of Karachentsev et al. (2007)
near M106 are common in our list except for the dwarf galaxy, 
d1217+4703 that is found to be located in one of the gaps.
This is consistent with the expected number of missing satellite 
galaxies due to the gaps between
CCDs in the present study. We do not include this satellite galaxy,
d1217+4703 for the subsequent analysis because we do not have its morphology 
and photometric properties.  

\begin{figure}
 \epsfig{figure=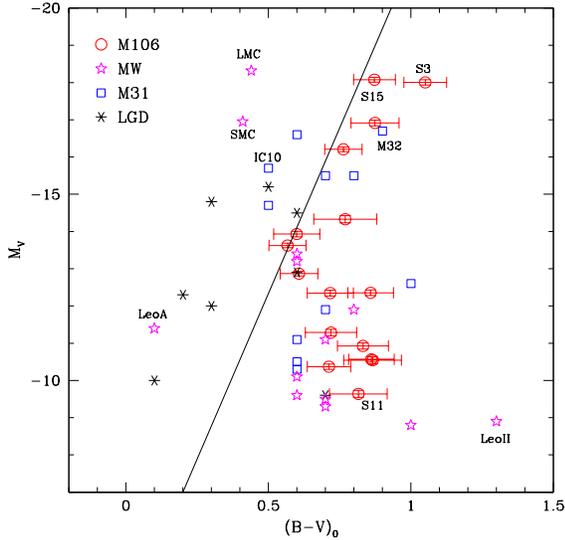, height=0.480\textwidth, width=0.480\textwidth}
 \vspace{3mm}
 \caption{
  $(B-V)_0$ versus $M_V$ diagram of the satellite galaxies of
  M106 (open circles with error bars). As a comparison
  the data for the Local Group dwarf galaxies are displayed with other symbols. The solid line
  represents a boundary to separate spirals or irregulars from dSph or Ellipticals \citep{mat98}.}
 \label{fig_galcmd}
\end{figure}

\begin{figure}
 \epsfig{figure=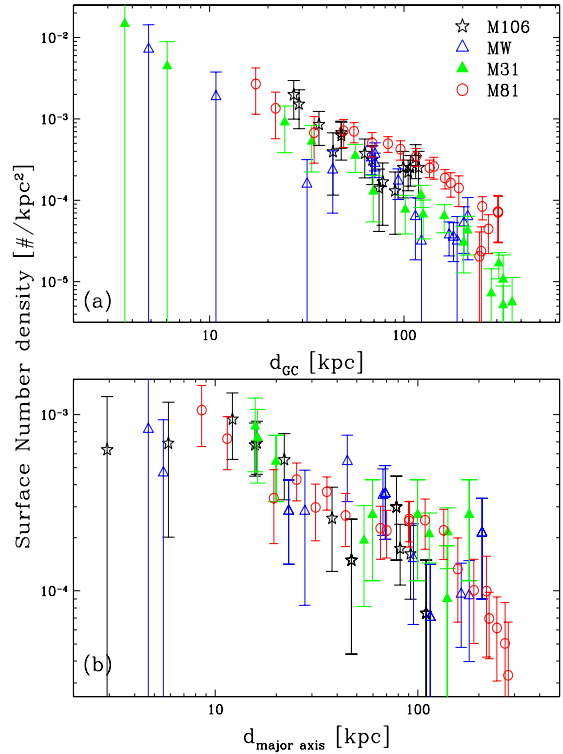, height=0.680\textwidth, width=0.480\textwidth}
 \vspace{3mm}
 \caption{
  Spatial distribution of the satellite galaxies of M106
  (star symbols), MW (open triangles), M31 (filled triangles),
  and the M81 group (open circles) against
  (a) sky-projected radial distance to the centre of each galaxy,
  and (b) sky-projected distance to the major axis
  of each galaxy. To mimic the data-set of M106 we transform
  the galacto-centric Cartesian coordinates of
  the satellite galaxies of Milky Way, M31, and the M81 group
  assuming an inclination angle of $67\degr$ for M106.}
 \label{fig_geom_comp}
\end{figure}

It is more difficult to estimate the number of missing 
satellite galaxies not located in these gaps.
One of the causes that make it difficult to identify the 
faint satellite galaxies of M106 is the observational 
limits; poor spatial resolution and
shallow photometric depth.
Identification of satellite galaxies of M106 based on the present 
observations is mainly dependent on surface brightness rather than
the total magnitude. Two satellite galaxies (S11 and S14) 
have low central surface brightness
($\mu_{0,V} \ge 25$ mag arcsec$^{-2}$). Therefore we conclude that 
satellite galaxies with extremely faint surface brightness may be 
missing in the present list of satellite galaxies of M106. 
This is more severe for the galaxies located in 
F3 because the measured sky brightness of F3 is much brighter than 
F1 and F2.

We discuss qualitatively the limit of our detection method by comparing 
our results th those 
of \citet{chi09} because their observations were also performed 
using CFHT MegaCam. \citet{chi09} found 22 new dwarf galaxies in the
M81 group using a larger field ($65$ square degree) MegaCam 
survey of the M81 group, resulting in a total of
44 galaxies in the M81 group including M81. To find the satellite galaxies they chose two different
methods; visual inspection by two of three authors, and a 2-point correlation auto-detection routine.
They detected 17 out of 22 new candidates of the M81 group by visual inspection of the 
MegaCam images. To find the satellite galaxies missed in the visual inspection they 
applied a 2-point correlation
auto-detection routine for the photometric catalogue of point sources. The auto-detection method
calculates the 2-point correlation function for the point sources with $22<r'<27$ on scales of
$0.02-0.11$ kpc, then a spatial region with a higher correlation values might be a candidate
satellite galaxy. \citet{chi09} applied this method for their photometric 
catalogue
of the M81 group because they were able to detect a large number of TRGB  
stars in the dwarf galaxies of the M81 group.
Assuming $M_{I,TRGB} = -4.05$ and typical colours of TRGB of
$(V-I)_{TRGB} \sim 1.5$, and $(V-R)_{TRGB} \sim 0.8$ \citep{lee93}, 
$M_{r',TRGB}$ is detected at $\approx -3.1$, 
corresponding to
$r'_{TRGB} \sim 24.7$ at the distance of M81. 
They argued that since their observational limit ($r' \sim 25.0 - 25.5$) 
is slightly fainter than the TRGB brightness they detected a significant amount of TRGB
stars in the satellite galaxies of the M81 group. Auto 
detection method adds 5 more satellite galaxies
to the catalogue of satellite galaxies of the M81 group,
mostly galaxies with faint surface brightness.
 
To investigate the limit of our detection method
we compare the observational parameters of both MegaCam observations. 
Our MegaCam observations of M106 is rather
deeper (exposure times in $r$-band is $\sim 2$ times longer than $r'$-band
observations of the M81 group). The atmospheric seeings of the both 
observations are similar to $\sim 0.75$ arcsec, and both
observations were carried out at dark nights. Therefore, the photometric 
limiting magnitude of the
present observation is expected to be $\sim 0.75$ mag fainter than the 
MegaCam observations of the M81 group. However, because the distance to
M106 is $\sim 2$ times farther than the distance to M81 ($d \approx 3.6$Mpc),
TRGB is expected to be detected at $r \sim 26.2$ for our MegaCam survey, 
similar to or fainter than the magnitude limit of the present MegaCam
observation of M106. Therefore, we expect to detect no or an extremely 
small number of TRGB stars in satellite galaxies of M106. Since 2-point 
correlation auto-detection method needs a catalogue of points sources
in the satellite galaxies we are not able to apply this method for
our MegaCam observations. Our catalogue of satellite galaxies of M106 does not
include the galaxies with extremely faint central surface brightness, which
were detected in the M81 group (4 out of 22 galaxies with $\mu_o(r')>27$).
We might miss those extremely faint galaxies in the present catalogue of
satellite galaxies of M106.
 
\subsection{Comparison with LG and the M81 group}

MW and M31 in LG have a large number of satellites whose properties are 
well known. The M81 group is known to have 43 satellite galaxies of which 
about half are new findings \citep{chi09}.
We compare the properties of the satellite
galaxies of M106 with those of MW, M31, and the the M81 group.
Because the brightness of 16 satellite galaxies of M106 is much brighter
than those of the extremely faint galaxies of MW, we only 
consider the brighter ($M_V\le-9$) satellite galaxies of MW
(samples in Mateo 1998; van den Bergh 2000; Metz et al. 2007).

We show the cumulative luminosity function (LF) of satellite galaxies of 
M106 compared to 
those of the satellite galaxies in LG and the M81 group in Fig. 7.
For M106 and the M81 group we show the $r$-band and $r'$-band cumulative LFs
while we display $V$-band cumulative LF for LG galaxies.
We exclude the error bars of data points in Fig. 7 for the clear comparison.
Then we fit the observed cumulative LFs to cumulative Schechter LFs 
(solid lines). To compare the LFs of
satellite galaxies we do not include the host galaxies of each system 
(MW, M31, M81, and M106).
For the M81 group, we also exclude the faintest sample from the fit to
the Schechter LF, which is significantly fainter than the majority of
satellite galaxies of the M81 group. The faint end slopes of 
Schechter LFs for M106, LG, and the M81 group are
$-1.19^{+0.03}_{-0.06}$, $-1.06^{+0.03}_{-0.03}$, 
and $-1.29^{+0.07}_{-0.03}$, respectively.
The faint end slope of the M81 group is basically  identical to that 
of \citet{chi09}.
The faint end slope of Schechter LF of M106 is slightly steeper than that 
of LG but flatter by $\approx 0.1$ compared to the M81 group.
The cumulative LF of the M81 group rises suddenly at $M_{r'} \approx -13$ due 
to the inclusion of a large number of satellite galaxies found in MegaCam 
survey \citep{chi09}. A significantly larger number of satellite galaxies
of the M81 group compared to M106 might be due to inclusion of some artifacts
and/or background galaxies as pointed out by \citet{chi09} (See the comments
of their Table 1). They performed spectroscopic confirmation of only
3 BCD candidates among 22 new satellite candidates, which turned out to
be the members of the M81 group. The cumulative LFs of LG and M106 satellite 
galaxies are represented well by the cumulative Schechter function, 
while the cumulative Schechter LF does not represent well the
observed cumulative LF of the M81 group as shown in Fig. 7.
The faint end slope of cumulative LF of
M106 is much flatter than the expected faint end slope by 
$\Lambda$CDM model of $-1.8$ (Trentham \& Tully, 2002).

In Fig. 8, we display $(B-V)_0-M_V$ colour-magnitude diagram of 16 
satellite galaxies in M106 (open circles with error bars).
We transform SDSS $g$ and $r$ magnitudes to Johnson $B$ and $V$ magnitudes using the
transformation relation by \citet{cho08}.
Since \citet{chi09} have observed the M81 group using only $r'$-band, we  
show only the colour-magnitude distribution of M106 and LG satellites.
The distribution in the colour-magnitude diagram (CMD)
for LG dwarf galaxies associated with neither MW nor M31 are shown as a comparison.
The solid line is a boundary based on observations to divide the dwarf galaxies
into early type galaxies of dSph or ellipticals (redder of the solid line) 
and late type galaxies of spirals or irregulars
(Mateo 1998). The locations of dwarf satellite galaxies of
M106 in the CMD coincide roughly with early type galaxies of M31 and MW. 
However the CMD distribution of LG dwarf galaxies not related with
large spiral galaxies is quite different
from that of M106 satellites. 

The mean colour of
dwarf satellite galaxies of M106 is estimated to be $<(B-V)_0>\approx0.73$,
consistent with that of the dwarf satellite galaxies of MW and M31.
Interestingly two brightest satellite
galaxies of M106 (S3 and S15) have much redder $(B-V)$ colour than
the dwarf galaxies of MW with similar luminosity. This difference is
related with the morphological difference between them. The brightest
satellite galaxy of M106 is an early type spiral galaxy, while that of MW is
a late type spiral. \citet{jam08} reported that it is quite rare to 
find star forming late
type spiral galaxies as a satellite galaxy of a giant spiral galaxy like MW.
However, there is a high probability of morphology conformity between
host and satellite galaxies \citep{ann08} and similar conformity is present
between a target galaxy and its companion, especially when the companion
is close enough to be located inside the virial radius of the target
galaxy \citep{park08}.

\begin{figure}
 \epsfig{figure=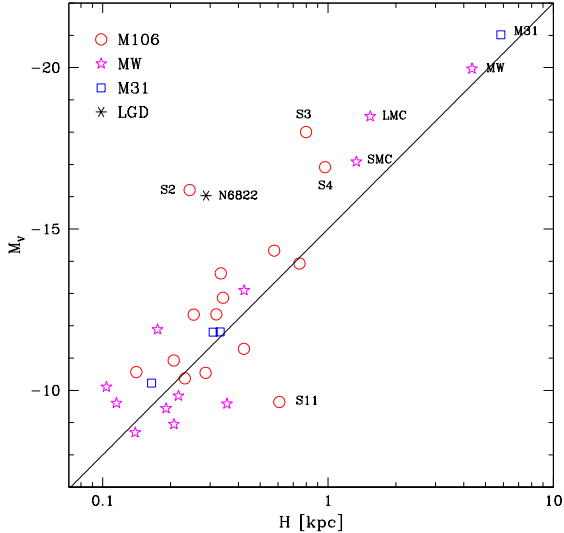, height=0.480\textwidth, width=0.480\textwidth}
 \caption{
  Distribution of dwarf galaxies of M106 and LG in
  disk scale length ($H$) versus $M_V$.
  For dwarf galaxies of LG we include dIrr and dSph.
  The solid line represents the relation for the
  dSph galaxies in LG \citep{bergh00}.}
 \label{fig_disk_comp}
\end{figure}

The spatial distribution of the satellite galaxies in M106 shows two 
distinct features; (1) galacto-centric concentration toward
M106 and (2) preferred concentration toward the major axis of M106. 
We compare the spatial distribution of the satellites of M106 to
those of the dwarf satellite galaxies in LG and the M81 group and display 
the results in 
Fig 9. The galacto-centric Cartesian coordinates of the satellite
galaxies of Milky Way and M31 are from Metz et al. (2007) and 
Koch \& Grebel (2006), 
respectively. To mimic the sky-projected
two dimensional distribution of M106, we transform their Cartesian 
coordinates by rotating along the semi-minor axis of the disk. 
We determine the inclination angle of M106 using the known major and 
minor-axis radii assuming zero-thickness of the disk
of M106 following Ma et al. (1997), obtaining $\sim 67\degr$.
Due to the limited survey area of the present study, the distribution of 
the satellite
galaxies of M106 is truncated at $\sim 130$kpc, corresponding to $\sim1\degr$. 

The surface number density profiles against
the sky-projected galacto-centric distance (Fig. 9(a)) are very similar to one
another among four satellite systems with an indication of slightly
higher number density of the satellite galaxies of the M81 group compared 
to MW, M31, and M106. The surface number density of the M81 group
is higher by a factor of $\sim 2.5$ than that of LG satellite galaxies for 
for the radial ranges of $50\sim200$kpc. We fit the radial surface number
density profiles of the satellites of MW, M31, and the M81 group with the 
power-law, obtaining the power indices of $-1.29\pm0.14$, $-1.36\pm0.17$,
and $-1.62\pm0.12$, respectively. For the brighter sample of the M81 group
with $M_{r'}<-10$, we obtain the power index of $-1.24\pm0.18$.
The radial surface number density profiles of the satellites of MW, M31, 
and the M81 group decrease gradually roughly following
the power-law with varying power indices. On the contrary, the 
radial distribution of satellites of M106 shows some hint of flat 
distribution for the outer region ($d_{M106}>\sim100$kpc).
M106 is expected to be located in rather denser environment than MW, M31, and
M81, this might affect the distribution of satellite galaxies of M106
compared to those of  MW, M31. and the M81 group.
However, since the number density profile of M106 satellites is
only available for the radial ranges of $d_{M106}<130$kpc, we are not able 
to conclude that if the outer distribution of M106 satellites follows
the uniform distribution. To reconcile this issue, it is need to extend the 
survey area to cover the whole dynamical boundary of M106.
We show the surface number density of the satellite galaxies of 
M106, MW, M31, and the M81 group with increasing projected distance from
the semi-major axis of each galaxy in Fig. 9(b). The distributions of the 
satellite galaxies of these galaxies are similar within the uncertainties.
We also compute the number of satellite galaxies of M106 by assuming that
the flat distribution in Fig. 5(a) ($d_{M106}>45$ arcmin) is mainly due to
background galaxies. We find that the number of satellite galaxies located 
between 
$10 - 45$ arcmin from M106 is $\sim8$ by assuming flat background distribution.
This number might be the lower limit of the actual satellite galaxies of M106,
corresponding to some half of candidate satellite galaxies found in this study.

To estimate the number of satellite galaxies of M106 located 
outside of the surveyed region ($d_{M106}>130$kpc)
of M106, we count the number of satellite galaxies outward of this
distance of MW, M31, and the M81 group by assuming that the virial mass
of each system is similar to each other. We find the numbers
4($\sim30\%$), 8($\sim50\%$), and 18($\sim40\%$) for MW, M31, and the M81 group,
respectively, with an average of 40\% of total satellite galaxies in these 
galaxies.
This simply implies that there might be $\sim 10$ satellite galaxies with
$M_r<-10$ in the outer region of M106. This estimation might be an upper
limit because the virial mass of M106 might be somewhat smaller than MW, M31, 
and the M81 group and the
catalogue of the satellite galaxies of the M81 group includes some very faint
galaxies down to $M_{r'}\sim-7$.
We compute the expected number of satellite galaxies of M106 located 
beyond the surveyed region of M106 ($d_{M106}>130$kpc)
by extrapolating the measured power-law slope of $-2.12$, 
resulting in N($>130$kpc) $\sim 6$.
This number is quite similar to that of M31, but significantly lower than those of
MW ($\sim11$), and the M81 group($\sim21$). 
Thus, to find the expected number of satellite galaxies in M106
we need to 
increase the survey area by a factor of $\sim 7$.

We show the distribution of dwarf galaxies of M106 in $M_V$ versus disk 
scale length ($H$) diagram in Fig. 10 (open circles).
For comparison the distribution of LG dwarf 
galaxies are shown with different symbols.
We do not include the data for the M81 group because \citet{chi09} presented
only the effective radii of the satellite galaxies.
The scale length data for LG galaxies including 
spirals, dIrr, and dSph are from \citet{bergh00}.
Most of dwarf galaxies of M106 are scattered in the domain of 
the distribution of dSph of LG. S11 classified as
dSph in this study has relatively large disk scale length 
than the other dSph galaxies with similar luminosity.
Late-type spiral satellite galaxies of M106 (S2, S3 and S4) are seen 
to lie well above the relation for dSph galaxies.
Overall the distribution of dwarf galaxies in M106 is very 
similar to that of dIrr and dSph of LG in
this diagram.

\section{Summary and Conclusion}

We present the result of a wide-area survey for satellite galaxies around M106.
Our survey is based on the large field deep $g$ and $r$ images obtained using 
the CFHT MegaCam
so that we could perform a systematic searching for satellite galaxies of M106. 
We summarize the main conclusions of the present study as follows.

\noindent
1. Based on $1.7\degr \times 2\degr$ deep observations of M106, we found
16 candidate satellite galaxies of M106, located as far as 
$\sim 130$kpc from the centre of M106. Four of
them are new findings. Due to the limited exposure times and
the large atmospheric seeing we are able to detect only a small number
of TRGB stars in the satellite galaxies of M106.
Our survey area might be significantly smaller than the physical
boundary of M106 dark halo. Therefore we might miss several satellite 
galaxies outside of $\sim130$kpc from M106.

\noindent
2. A satellite system of M106 shows a clear signature of 
central concentration toward
M106 and is more concentrated toward the semi-major axis of M106. 
We fit the inner region of the radial
surface number density profiles with a power-law, 
resulting a power index of $-2.1\pm0.5$, 
similar to that of the isothermal distribution.

\noindent
3. The surface brightnesses of 15 satellites galaxies are found to fit well 
with an exponential profile while the remaining brightest satellite galaxy 
of M106 (S15) 
shows a radial profile of the typical early type galaxy.

\noindent
4. The luminosity of 16 satellite galaxies spans a large 
range ($\Delta M_V \approx 7.6$ mag), but $(B-V)$ colours have
a mean value of 0.73 with a small dispersion.
The luminosity function of the satellite galaxies of M106 is 
well represented by the Schechter function
with a faint end slope of $-1.19$.

\noindent
5. The spatial distribution of the satellite galaxies of M106, Milky Way, 
M31, and the M81 group is found to be
similar to one another.

\section*{Acknowledgments}

Authors thank the anonymous referee for constructive comments and suggestions.
EK was supported in part by the National Research Foundation of Korea 
to the Center for Galaxy Evolution Research and by the second phase of
the Brain Korea 21 Program in 2010.
MK was supported by Basic Science Research Program through the
National Research Foundation of Korea (NRF)
funded by the Ministry of Education, Science and Technology (2010-0007713).
NH was supported in part by Grant-in-Aid for JSPS Fellow No. 20-08325.
MGL was supported by Mid-career Researcher Program
through NRF grant funded by the MEST (No.2010-0013875).
HBA thanks to ARCSEC for the support from the National Research 
Foundation of Korea grant funded by the Korea government (2010-001308).

\end{document}